\documentclass[conference, 9pt]{IEEEtran}

\usepackage[utf8]{inputenc} 
\usepackage[T1]{fontenc}
\usepackage{url}              

\usepackage[cmex10]{amsmath}  
\usepackage{mleftright}       
\mleftright                   

\usepackage{amsthm, amsmath, amssymb, amsfonts, aliascnt, balance, mathtools,graphicx,textcomp,xcolor, xpatch, pgfplots, tikz, forest, epstopdf, pdftexcmds, floatrow, subfiles, csquotes, enumitem} 

\usepackage{flowchart, pgf, bmpsize, adjustbox, pdftexcmds, float, array, arydshln, multirow, rotating, bm, amsbsy, pst-node, lipsum, tikzscale, tkz-euclide, mathrsfs, booktabs, mathtools, makecell, float, romannum}

\usetikzlibrary{positioning, arrows.meta, shapes, backgrounds}
\usepackage{booktabs}         
\usepackage[backend=bibtex,bibstyle=ieee,citestyle=numeric-comp,url=false,doi=false,isbn=false]{biblatex}

\usepackage[draft]{hyperref}
\hypersetup{
	colorlinks   = true, 
	urlcolor     = blue, 
	linkcolor    = blue, 
	citecolor   = red 
}

\newcounter{cases}
\newcounter{subcases}[cases]

\makeatletter
\xpatchcmd{\proof}{\topsep6\p@\@plus6\p@\relax}{}{}{}
\makeatother

\newtheorem*{lemma*}{Lemma}
\newtheorem*{theorem*}{Theorem}

\DeclarePairedDelimiterX\mathset[2]{\lbrace}{\rbrace}{#1 : #2}

\newcommand{\bfd}{{\boldsymbol d}}

\newcommand{\bfh}{{\boldsymbol h}}

\newcommand{\bfs}{{\boldsymbol s}}

\newcommand{\bfu}{{\boldsymbol u}}
\newcommand{\bfv}{{\boldsymbol v}}

\newcommand{\bfx}{{\boldsymbol x}}
\newcommand{\bfy}{{\boldsymbol y}}

\newcommand{\bfH}{{\mathbf H}}

\newcommand{\bfzero}{{\mathbf 0}}

\newcommand{\cB}{\mathcal{B}}
\newcommand{\cC}{\mathcal{C}}

\newcommand{\cF}{\mathcal{F}}

\usepackage[nameinlink]{cleveref}
\crefname{section}{Section}{Sections}
\crefname{subsection}{Section}{Sections}
\crefname{equation}{Eq.}{Equations}
\crefname{enumi}{part}{parts}
\crefname{table}{Table}{Tables}
\crefname{figure}{Figure}{Figures}
\crefname{algocf}{Algorithm}{Algorithms}

\newaliascnt{conjecture}{theorem}

\aliascntresetthe{conjecture}
\crefname{conjecture}{Conjecture}{Conjectures}

\newaliascnt{question}{theorem}

\aliascntresetthe{question}
\crefname{question}{Question}{Questions}

\newaliascnt{example}{theorem}
\newtheorem{example}[example]{Example}
\aliascntresetthe{example}
\crefname{example}{Example}{Examples}

\tikzset{
	dot/.style={circle,draw,inner sep=0.5,fill=black},
}

\newcommand{\pin}{P_{\mathrm{i}}}
\newcommand{\pd}{P_{\mathrm{d}}}
\newcommand{\ps}{P_{\mathrm{s}}}
\newcommand{\pt}{P_{\mathrm{t}}}

\pgfplotsset{compat=newest}
\pgfplotsset{plot coordinates/math parser=false}
\usepgfplotslibrary{colormaps} 
\pgfplotsset{compat=1.9}
\pgfplotsset{
	colormap={bright}{rgb255=(0,180,0) rgb255=(2,74,255)
		rgb255=(230,0,0) rgb255=(0,204,238) rgb255=(0,0,0) rgb255=(245,0,245) rgb255=(190,225,0) rgb255=(147,213,114)  rgb255=(255,113,26) rgb255=(255,21,181)} 
}

\newlength\figureheight
\newlength\figurewidth

\pgfplotsset{fano/.style={
		blue,             
		solid,            
		mark=square*,     
		mark options={fill=blue} 
}}

\pgfplotsset{stack/.style={
		red,             
		mark options={fill=red} 
}}
\pgfplotsset{bistack/.style={
		blue,             
		mark options={fill=blue} 
}}
\pgfplotsset{bcjr/.style={
		green!20!black,             
		mark options={fill=green!20!black} 
}}

\bibliography{resources/literature.bib}

\IEEEoverridecommandlockouts

\begin{document}

\title{Sequential Decoding of Multiple Traces Over the Syndrome Trellis for Synchronization Errors  

\vspace{-2mm}

\thanks{Funded by the European Union (DiDAX, 101115134). This work was partially supported by the Swedish Research Council (VR) under grant 2020-03687. }}

\author{%
  \IEEEauthorblockN{Anisha~Banerjee\IEEEauthorrefmark{1}, Lorenz~Welter\IEEEauthorrefmark{1}, Alexandre~Graell~i~Amat\IEEEauthorrefmark{2},                     
    Antonia~Wachter-Zeh\IEEEauthorrefmark{1},
    and Eirik~Rosnes\IEEEauthorrefmark{3}
    }
  \IEEEauthorblockA{\IEEEauthorrefmark{1}%
                    Institute for Communications Engineering, Technical University of Munich (TUM), Munich, Germany}
  \IEEEauthorblockA{\IEEEauthorrefmark{2}Department of Electrical Engineering, Chalmers University of Technology, SE-41296 Gothenburg, Sweden}
  \IEEEauthorblockA{\IEEEauthorrefmark{3}Simula UiB, N-5006 Bergen, Norway}
    \IEEEauthorblockA{Email: anisha.banerjee@tum.de, lorenz.welter@tum.de, alexandre.graell@chalmer.se, \\ antonia.wachter-zeh@tum.de, eirikrosnes@simula.no }
      \\[-3.6ex]

      \vspace{-5mm}

}

\maketitle

\begin{abstract}
Standard decoding approaches for convolutional codes, such as the Viterbi and BCJR algorithms, entail significant complexity when correcting synchronization errors. The situation worsens when multiple received sequences should be jointly decoded, as in DNA storage. Previous work has attempted to address this via separate-BCJR decoding, i.e., combining the results of decoding each received sequence separately. Another attempt to reduce complexity adapted sequential decoders for use over channels with insertion and deletion errors. However, these  decoding alternatives remain prohibitively expensive for high-rate convolutional codes. To address this,  we adapt sequential decoders to decode multiple received sequences jointly over the \textit{syndrome trellis}. For the short blocklength regime, this decoding strategy can outperform separate-BCJR decoding under certain channel conditions, in addition to reducing decoding complexity. To mitigate the occurrence of a decoding timeout, formally called erasure, we also extend this approach to work bidirectionally, i.e., deploying two independent stack decoders that simultaneously operate in the forward and backward directions.
\end{abstract}


%
\IEEEpeerreviewmaketitle

\section{Introduction}

Synchronization errors, namely insertions and deletions, affect several networking and data storage channels \cite{sklarDigitalCommunicationsFundamentals2021, kinnimentSynchronizationArbitrationDigital2007, schouhamerimminkCodesMassData2004, heckelCharacterizationDNAData2019, yazdiDNABasedStorageTrends2015}. Such errors involve the loss of symbols during transmission or the addition of some spurious symbols into the received stream, respectively, and usually arise either from intrinsic characteristics of the channel, e.g., in DNA data storage, or due to external causes such as imperfect synchronization between the transmitter and the receiver. A wide variety of error-correcting codes \cite{mercier_survey_2010, Levenshtein65, calabiGeneralResultsCoding1969, briffa_timevarying_2014, coumou_insertion_2008} have been investigated to improve transmission reliability over channels with synchronization errors. This work focuses on convolutional codes, specifically their decoding algorithms for channels with synchronization errors. Prior work \cite{mansour_convolutional_2010,buttigiegImprovedBitError2015} proposed new trellis structures that facilitate the use of Viterbi and maximum \textrm{a posteriori}  decoders for correcting insertions and deletions, in addition to substitution errors. However, these trellises grow exponentially with the memory of the convolutional code, and the maximum number of insertions and deletions considered, resulting in high memory and decoding complexity. The increase in complexity is particularly significant when multiple sequences are decoded jointly \cite{maaroufConcatenatedCodesMultiple2023, sakogawaSymbolwiseMAPEstimation2020}. To mitigate this issue, the solution proposed in \cite{maaroufConcatenatedCodesMultiple2023} involves decoding each received sequence independently before combining their respective symbol-wise a posteriori probabilities. However, this incurs a loss in achievable rate compared to joint decoding of multiple sequences \cite{maaroufConcatenatedCodesMultiple2023}. Another efficient albeit suboptimal alternative is sequential decoding \cite{wozencraftSequentialDecodingReliable1957, jelinekFastSequentialDecoding1969, fanoHeuristicDiscussionProbabilistic1963, banerjeeSequentialDecodingMultiple2024}, which operates on a tree representation of the code instead of its trellis and only examines codewords (paths in the tree) that appear to be promising. This method has a complexity that scales linearly with the code length and does not depend on the memory of the code. This remedy is also not perfect since the complexity of sequential decoding grows exponentially with the dimension of the code and the number of received sequences \cite{banerjeeSequentialDecodingMultiple2024}. Furthermore, both approaches render the use of high-rate convolutional codes impractical due to the explosion in the number of outgoing edges per node in the trellis. 

In this work, we address these drawbacks by turning to an alternative representation of convolutional codes based on the parity-check matrix, namely the \emph{syndrome trellis (tree)} \cite{wolfEfficientMaximumLikelihood1978, bahlOptimalDecodingLinear1974, sidorenkoDecodingConvolutionalCodes1994}. Specifically, we adapt sequential decoders to operate on the syndrome tree by refining the decoding metric to accommodate the tree's irregular nature. This strategy makes the complexity of sequential decoding independent of the dimension of the code in low-noise environments, hence enabling the use of high-rate codes. We also reduce the probability of a decoding timeout, formally called  erasure, by extending the stack algorithm to its bidirectional counterpart \cite{kallelBidirectionalSequentialDecoding1997}. 
We show that for short blocklength codes, jointly decoding multiple sequences over the syndrome tree with the stack algorithm outperforms separate-BCJR decoding under specific noise and code rate regimes.

\section{Preliminaries}
\subsection{Channel Model} \label{subsec::channel}

\begin{figure}[t]
	\centering
	\begin{tikzpicture}[->,>={Stealth},auto,node distance=3.5cm,scale=0.86, every node/.style={scale=0.8}]
	\node[name=a, circle, inner sep=0pt, draw=black,fill=white, minimum size=1cm] at (0,-2){$x_{i}$};
	\node[name=b, circle, inner sep=0pt, draw=black,fill=white, minimum size=1cm] at (6.0,-2){$x_{i+1}$};
	\node[anchor=east,xshift=-15pt] at (a) {$\cdots$};
	\node[anchor=east,xshift=40pt] at (b) {$\cdots$};
	\path[]
	(a) edge[bend right=35, dashed] node [midway, below, yshift=-.7mm] {$\pd$} node[midway, above] {Delete}(b)
	(a) edge[bend left=35] node [midway, above, right, yshift=3mm, xshift=-18.5mm,name=tr] {Transmit with no error} node[midway, below] {$\pt(1-\ps)$} (b);
	\draw (a) edge[loop above, out=120, in=60,min distance=20mm] node {Insert} node[midway, left, xshift=-5mm, yshift=-8mm] {$\pin$}(a);
	\draw (a) ->node [above] {Transmit with substitution} node [below] {$\pt\ps$}  (b);
	\end{tikzpicture}
	\caption{Allowed transitions in the state machine model for the insertion, deletion, and substitution channel \cite{daveyReliableCommunicationChannels2001}.}
	\label{fig::channelDavey}
\end{figure}
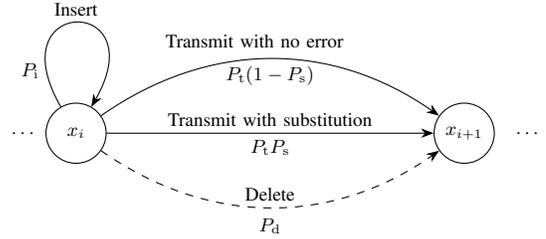

As in \cite{daveyReliableCommunicationChannels2001, bahlOptimalDecodingLinear1974}, we model the channel as a finite-state machine specified by parameters $\pin$, $\pd$, and $\ps$, representing the insertion, deletion, and substitution probabilities, respectively. Say a sequence $\bfx = (x_1, \ldots, x_L) \in \{0,1\}^L$ is awaiting transmission. As illustrated in Fig.~\ref{fig::channelDavey}, for each input bit $x_i$, one of the following events may occur: \romannum{1}) with probability $\pin$, a random bit is inserted into the received stream, and $x_i$ remains in the transmission queue; \romannum{2}) $x_i$ is deleted with probability $\pd$; \romannum{3}) $x_i$ is received either correctly or with a substitution, with probability $\pt(1-\ps)$ and $\pt\ps$, respectively. Here, $\pt = 1 - \pin - \pd$ represents the transmission probability.

\subsection{Convolutional Codes and the Syndrome Trellis}

An $[n,k]$ convolutional code can be defined by an $(n-k) \times n$ polynomial parity-check matrix $\bfH(D)$. An $[N,K]$ terminated convolutional code $\cC$ formed from this $[n,k]$ convolutional code, can be defined by a binary parity-check matrix $\bfH = (\bfh_1, \ldots, \bfh_N)$ so that each codeword $\bfx = (x_1, \ldots, x_N) \in \cC$ satisfies $\sum_{i=1}^{N}x_i\bfh_i= \bfzero$, where $N$ is the length of the codeword. This motivates the concept of a \textit{syndrome trellis}, which comprises $N+1$ levels, each containing multiple nodes, each representing a specific \emph{syndrome state}. Each codeword $\bfx \in \cC$ traces a path in the trellis that starts from $\bfs_0(\bfx) = \bfzero$ and subsequently passes through the syndrome states $\bfs_l(\bfx) = x_1\bfh_1 + \cdots + x_l\bfh_l$ at levels $l = 1, \ldots, N$.


\begin{example} \label{eg::cc}
    Consider the $[3,2]$ convolutional code defined by the parity-check matrix $\bfH(D) = \begin{bmatrix}
             1+ D & 1+D^2 & 1+D+D^2
        \end{bmatrix}$.
    The associated syndrome trellis, shown in Fig.~\ref{fig::cc32} for $N=9$, is built using the corresponding binary parity-check matrix.
    \begin{equation*}
        \bfH = \begin{bmatrix}
            1 & 1 & 1 & & &  & & & \\
            1 & 0 & 1 & 1 & 1 & 1 &  & & \\
            0 & 1 & 1 & 1 & 0 & 1 & 1  & 1 & 1\\
              &   &   & 0 & 1 & 1 & 1  & 0 & 1\\
              &   &   &   &   &   & 0  & 1 & 1\\
        \end{bmatrix}\,.
    \end{equation*}
    
    
\end{example}


\begin{figure}[t]
    \centering
    \scalebox{1.1}{\begin{tikzpicture}[long dash/.style={dash pattern=on 10pt off 2pt}, short dash/.style={ dash pattern=on 3pt off 2.4pt}]
    \tikzstyle{tiny dot}=[circle, fill, inner sep=0.5pt]
    \newlength{\vsep}
    \newlength{\hsep}
    \setlength{\vsep}{0.35cm}
    \setlength{\hsep}{0.67cm}

    \foreach \j/\col in {-3/green, -2/green, 0/green, 1/green, -1/blue, 2/blue, 3/red, 4/red, 5/red}{
        \fill[fill=\col!20, opacity=0.4] ($(\j*\hsep, 0)$) rectangle ($(\j*\hsep + \hsep, -7*\vsep)$);
    }

    \foreach \i in {0,1,2,3,4,5,6,7}{
        \foreach \j in {-3,-2,-1,0,1,2,3,4,5,6}{
            \node[draw, tiny dot] (x\i\j) at ($(\j*\hsep, -\i*\vsep)$){};
    }
    \foreach \j/\si/\se in {-3/0/0,-2/0/0,-2/3/3,-1/0/0, -1/6/3}{
        \draw[short dash] ($(\j*\hsep, -\si*\vsep)$) -- ($(\j*\hsep + \hsep, -\se*\vsep)$);
    }
    \foreach \j/\si/\se in {-3/0/3,-2/3/6, -2/0/5, -1/3/2, -1/5/1}{
                \draw[solid] ($(\j*\hsep, -\si*\vsep)$) -- ($(\j*\hsep + \hsep, -\se*\vsep)$);
    }
    \foreach \i in {0,1,2,3}{
        \foreach \j in {0,1}{
            \draw[short dash] ($(\j*\hsep, -\i*\vsep)$) -- ($(\j*\hsep, -\i*\vsep) + (\hsep,0)$);
        }
    }
    \foreach \si/\se in {0/0, 2/1, 4/2, 6/3}{
        \draw[short dash] ($(2*\hsep, -\si*\vsep)$) -- ($(3*\hsep, -\se*\vsep)$);
    }        
    \foreach \ji/\si/\se in {0/0/3, 0/1/2, 0/2/1, 0/3/0, 1/0/5, 1/1/4, 1/2/7, 1/3/6, 2/1/3, 2/3/2, 2/5/1, 2/7/0}{
        \draw[solid] ($(\ji*\hsep, -\si*\vsep)$) -- ($(\ji*\hsep +\hsep, -\se*\vsep)$);
    }
    \foreach \j/\si/\se in {3/0/0, 4/0/0, 5/0/0, 3/2/2}{
        \draw[short dash] ($(\j*\hsep, -\si*\vsep)$) -- ($(\j*\hsep + \hsep, -\se*\vsep)$);
    }
    \foreach \j/\si/\se in {3/1/2, 3/3/0, 4/2/7, 5/7/0}{
        \draw[solid] ($(\j*\hsep, -\si*\vsep)$) -- ($(\j*\hsep + \hsep, -\se*\vsep)$);
    }

    \draw[dotted] ($(0*\hsep, 0) + (-3pt, 3pt)$) rectangle ($(3*\hsep, -7*\vsep) + (3pt, -3pt)$);

    \foreach \j/\lev in {-3/0,-2/1,-1/2,0/3,1/4,2/5,3/6,4/7,5/8,6/9}{
        \node[font=\scriptsize] (y0\j) at ($(\j*\hsep, 6.5pt)$) {$\lev$};
    }

    }
    \end{tikzpicture}}
    \caption{Syndrome trellis of the $[3,2]$ convolutional code of length $N=9$ in Example~\ref{eg::cc}. Dashed and solid edges emerging from a node at level $l$ represent codeword bits $x_{l+1}=0$ and $x_{l+1}=1$, respectively. For longer codes, the trellis consists of successive repetitions of the block within the dashed box. The final three stages show the termination phase. Levels $0, 1, 3, 4$ are information levels while levels $2, 5, 6,7, 8$ are parity levels. The green, blue,  and red stages indicate information bits, parity bits, and termination bits, respectively.
    } 
    \label{fig::cc32}
\end{figure}


Before its terminating phase, the syndrome trellis of an $[n,k]$ convolutional code consists of consecutive blocks, each composed of $n$ trellis sections, $k$ of which correspond to information bits, while the remaining $n-k$ sections correspond to parity bits. Naturally, each node at the start of an information trellis section has two emanating edges, for $0$ and $1$, respectively, and is formally said to be at an \emph{information level}. The nodes at the start of a parity trellis section are said to be at a \emph{parity level} and only produce one edge. To encode an information vector using the trellis, one starts at the syndrome state $\bfs=\bfzero$ and chooses the edge corresponding to the next unencoded information bit when at an information level. When at a parity level or in the terminating phase, one follows the only outgoing edge. The final codeword is the concatenation of the edge labels along the path traced in the trellis. The syndrome trellis for the $[3,2]$ code in Example~\ref{eg::cc} is depicted in Fig.~\ref{fig::cc32}.


\begin{example} \label{eg::cc_encode}
    Considering the $[3,2]$ convolutional code from Example~\ref{eg::cc}, we observe from Fig.~\ref{fig::cc32} that the information vector $\bfu = (1, 0, 1, 1)$ maps to the codeword $\bfx = (1, 0, 1, 1, 1, 0, 0, 1, 1)$.
\end{example}

\subsection{Joint  Channel and Code Tree} \label{subsec::channel-code-tree}

The presence of insertion and deletion errors suggests that the channel has infinite memory. Thus, to interpret the received vector as the output of a hidden Markov model (HMM), we use the concept of a drift variable as in \cite{buttigiegImprovedBitError2015}. More explicitly, we view the received vector as the output of an HMM wherein each hidden state is a pair of a syndrome state and a drift value. The \emph{drift} $d_i$ \cite{daveyReliableCommunicationChannels2001} refers to the difference between the number of insertion and deletion events that occurred until the $(i+1)$th bit has been enqueued for transmission. The sequential decoder, as discussed in Section~\ref{subsec::seq-dec} below, works on a tree representation of this HMM, an example of which is depicted in Fig.~\ref{fig::hmm_code}. 
To facilitate the joint decoding of $M$ received sequences, the HMM must be augmented such that every hidden state combines a syndrome state with $M$ drift variables \cite{maaroufConcatenatedCodesMultiple2023}, one for each of the received sequences.

\begin{figure}[t]
	\centering
	\scalebox{0.9}{
		\begin{forest}
			[ ,name=a,for tree={s sep=2.7pt,l sep=2.4cm, dot, grow=0}
			[, name=b1, edge label={node[below,midway,yshift=-3pt,font=\scriptsize]{}},
			[, name=c1, fill=none, draw=none, edge={draw=none}]
			[, name=c2, fill=none, draw=none, edge={draw=none}]
			[, name=c3, fill=none, draw=none, edge={draw=none}]
			[, name=c4, fill=none, draw=none, edge={dashed, draw=none}]
			[, name=c5, fill=none, draw=none, edge={dashed, draw=none}]
			[, name=c6, fill=none, draw=none, edge={dashed, draw=none}]]
			[, name=b2, edge label={node[below,midway,yshift=-1pt,font=\scriptsize]{}},
			[, name=c7, ]
			[, name=c8, 
            [,name=d1, edge={dashed},] 
            [,name=d2, edge={dashed},] 
            [,name=d3, edge={dashed},]]
			[, name=c9]
			[, name=c10, edge={dashed},]
			[, name=c11, edge={dashed}, 
            [,name=d4] [,name=d5] [,name=d6]]
			[, name=c12,  edge={dashed}]]
			[, name=b3, edge label={node[below,midway,yshift=1pt,font=\scriptsize]{}},
			[, name=c13, fill=none, draw=none, edge={draw=none}]
			[, name=c14, fill=none, draw=none, edge={draw=none}]
			[, name=c15, fill=none, draw=none, edge={draw=none}]
			[, name=c16, fill=none, draw=none, edge={dashed, draw=none}]
			[, name=c17, fill=none, draw=none, edge={dashed, draw=none}]
			[, name=c18, fill=none, draw=none, edge={dashed, draw=none}]]
			[, name=b4, edge={dashed},
			[, name=c19, fill=none, draw=none, edge={draw=none}]
			[, name=c20, fill=none, draw=none, edge={draw=none}]
			[, name=c21, fill=none, draw=none, edge={draw=none}]
			[, name=c22, fill=none, draw=none, edge={dashed, draw=none}]
			[, name=c23, fill=none, draw=none, edge={dashed, draw=none}]
			[, name=c24, fill=none, draw=none, edge={dashed, draw=none}]]
			[, name=b5, edge label={node[above,midway,yshift=1.5pt,font=\scriptsize]{}}, edge={dashed},
			[, name=c25]
			[, name=c26, [, name=d7] [, name=d8] [, name=d9]]
			[, name=c27 ]
			[, name=c28, edge={dashed}]
			[, name=c29, edge={dashed}, 
            [, name=d10, edge={dashed}] 
            [, name=d11, edge={dashed}] 
            [, name=d12, edge={dashed}]]
			[, name=c30, edge={dashed}]]
			[, name=b6, edge label={node[above,midway,yshift=3.5pt,font=\scriptsize]{}}, edge={dashed},
			[, name=c31, fill=none, draw=none, edge={draw=none}]
			[, name=c32, fill=none, draw=none, edge={draw=none}]
			[, name=c33, fill=none, draw=none, edge={draw=none}]
			[, name=c34, fill=none, draw=none, edge={dashed, draw=none}]
			[, name=c35, fill=none, draw=none, edge={dashed, draw=none}]
			[, name=c36, fill=none, draw=none, edge={dashed, draw=none}]]]
			\node[yshift=-0.28cm,xshift=-0.3cm,font=\scriptsize] at (a) {$(S_0, 0)$};
			\node[yshift=-0.25cm,font=\scriptsize] at (b1) {$(S_3, 1)$};
            \node[xshift=0.3cm,font=\scriptsize] at (b1) {$\cdots$};
			\node[yshift=-0.25cm,font=\scriptsize] at (b2) {$(S_3, 0)$};
			\node[yshift=-0.25cm,font=\scriptsize] at (b3) {$(S_3, -1)$};
            \node[xshift=0.3cm,font=\scriptsize] at (b3) {$\cdots$};
			\node[yshift=0.25cm,font=\scriptsize] at (b4) {$(S_0, 1)$};
            \node[xshift=0.3cm,font=\scriptsize] at (b4) {$\cdots$};
			\node[yshift=0.25cm,font=\scriptsize] at (b5) {$(S_0, 0)$};
			\node[yshift=0.25cm,font=\scriptsize] at (b6) {$(S_0, -1)$};
            \node[xshift=0.3cm,font=\scriptsize] at (b6) {$\cdots$};
			\node[yshift=-0.25cm,font=\scriptsize] at (c7) {$(S_6, 1)$};
            \node[yshift=0.35cm,font=\scriptsize] at (c11) {$(S_3, -1)$};
			\node[yshift=-0.25cm,font=\scriptsize] at (d1) {$(S_3, 1)$};
            \node[yshift=0.25cm,font=\scriptsize] at (d6) {$(S_2, -1)$};
            \node[yshift=-0.25cm,font=\scriptsize] at (c25) {$(S_5, 1)$};
            \node[yshift=0.25cm,font=\scriptsize] at (c30) {$(S_0, -1)$};
            \node[yshift=-0.25cm,font=\scriptsize] at (d7) {$(S_2, 1)$};
            \node[yshift=0.25cm,font=\scriptsize] at (d12) {$(S_0, -1)$};
            \foreach \i in {7, 9, 10, 12, 25, 27, 28, 30} {
                \node[xshift=0.3cm,font=\scriptsize] at (c\i) {$\cdots$};
            }
            \begin{scope}[on background layer]
                \fill[fill=green!20, opacity=0.4] (c36.center) rectangle ($(c1.center) + (-4.9cm, 0)$);
                \fill[fill=blue!20, opacity=0.4] (c36.center) rectangle ($(c1.center) + (2.5cm, 0)$);
            \end{scope}
	\end{forest}}
	\caption{Part of the joint channel and code tree of the $[3,2]$ convolutional code from Example~\ref{eg::cc}. Dashed and solid lines indicate outputs $0$ and $1$, respectively. Each node represents a pair of a syndrome state and a drift state $(\bfs_l, d_j)$. The notations $S_0, \ldots, S_6$ represent specific realizations of a syndrome state. For ease of illustration, we restrict the net drift per branch to $\{-1, 0, 1\}$. The green and blue stages indicate information bits and parity bits, respectively.}
	\label{fig::hmm_code}
\end{figure}

\subsection{Stack Algorithm} \label{subsec::seq-dec}

Like other variants of sequential decoding, the stack algorithm \cite{jelinekFastSequentialDecoding1969} works on a tree representation of the HMM described earlier and evaluates the nodes visited in this tree through a decoding metric (discussed in Section~\ref{subsec::branch-metric}). This metric quantifies how closely the received sequence resembles the (partial) codeword and the sequence of drift states represented by the unique path that connects the tree root to this node. It employs a stack to store the paths that have already been visited and aims to find the path in the tree with the highest metric by executing the following steps.
\begin{enumerate}
    \item Place the tree's root node (with metric zero) at the top of the empty stack.
    \item Remove the top node and compute the metrics of its immediate successors. Next, these successor nodes are inserted into the stack while maintaining the decreasing order of node metrics in the stack. If the stack reaches its maximum size, remove the node at the bottom of the stack to make room for new nodes.
    \item If the top node is at the end of the tree, stop and output its associated codeword. If a maximum number of iterations (of Step 2) is reached, go to Step 4. Otherwise, go to Step 2.
    \item Declare an erasure. Pick the top node in the stack, and output the corresponding (incomplete) codeword and randomly guess remaining undecided bits.
\end{enumerate}

\section{Decoder Metric}

\subsection{Node Metric Over the Syndrome Trellis} \label{subsec::branch-metric}

Let $\bfy_1^R=(y_1, \ldots, y_R)$ be a sequence received when transmitting a codeword  $\bfx$ of length $N$. For a node $\bfv_0^t$ at depth $t$ in the joint channel and code tree, the path from the root to the node $\bfv_0^t$ determines a sequence of syndrome states, $\bfs_0^t=(\bfs_0, \ldots, \bfs_t)$ and drift states, $\bfd_0^t=(d_0, \ldots, d_{t})$, where the initial drift is $d_0=0$. As in \cite{masseyVariableLengthCodesFano1972, banerjeeSequentialDecodingMultiple2024}, we define the metric 
\begin{align*}
    \mu(\bfv_0^t) &\triangleq \log P(\bfv_0^t, \bfy_1^R) - \log \sum_{\bfx \in \{0,1\}^N} P(\bfx, \bfy_1^R)\,, 
\end{align*}
which approximates the posterior probability of the path $\bfv_0^t$ given the sequence $\bfy_1^R$, i.e., $\log P(\bfv_0^t | \bfy_1^R)$.  Since $\bfv_0^t$ only accounts for the first $t+d_t$ symbols of $\bfy_1^R$, we compute $P(\bfv_0^t, \bfy_1^R)$ by marginalizing over all possible codeword tails that could lead to the reception of symbols $\bfy_{t+d_t+1}^R$, while the encoder traverses a specific sequence of syndrome states from time $t+1$ to $N$, say $\widetilde{\bfs}_{t+1}^N$, i.e.,
\begin{equation}
    P(\bfv_0^t, \bfy_1^R) \!
    = \! P(\bfy_1^{t+d_t}, \bfv_0^t) \!\! \sum_{\widetilde{\bfs}_{t+1}^N} \! P(\widetilde{\bfs}_{t+1}^N, \bfy_{t+d_t+1}^R | \bfs_t, d_t)\,. \label{eq::tail1}
\end{equation}
For ease of evaluation, we apply the following approximation,
\begin{align*}
    \sum_{\widetilde{\bfs}_{t+1}^N} \! P(\widetilde{\bfs}_{t+1}^N, \bfy_{t+d_t+1}^R | \bfs_t, d_t) &\approx \!\! \! \sum_{\widetilde{\bfx}_{t+2}^N \in \{0,1\}^{N - t - 1}} \!\!\!\! P(\widetilde{\bfx}_{t+2}^N, \bfy_{t+d_t+1}^R | d_t).
\end{align*}
Since the initial syndrome and drift states are always $\bfs_0=\bfzero$ and $d_0=0$, it follows from the Markov chain-like behavior of the sequence of syndrome states $\bfs_0^t$ and drift states $\bfd_0^t$ that 
\begin{align*}
    P(\bfy_1^{t+d_t}, \bfv_0^t) &=  \prod_{i=0}^{t-1} P(\bfs_{i+1}|\bfs_i) P(\bfy_{i+d_i+1}^{i+1+d_{i+1}}, d_{i+1}| \bfs_i^{i+1}, d_i)\,.
\end{align*}
Thus, the metric of the node $\bfv_0^t$ evaluates to
\begin{align}
    \mu(\bfv_0^t) \! &\approx \! \sum_{i=0}^{t-1 } \! \Big( \! \log P(\bfs_{i+1}|\bfs_i) \!+\! \log P(\bfy_{i+d_i+1}^{i+1+d_{i+1}}, d_{i+1}|\bfs_i^{i+1},d_i) \!\Big)  \nonumber \\
    &+ \log \!\!\! \sum_{\bfx \in \{0,1\}^{N - t - 1}} \!\!\! P(\bfx, \bfy_{t+d_t+1}^R| d_t)  -  \log \!\!\! \sum_{\bfx \in \{0,1\}^N} \!\!\! {P(\bfx, \bfy_1^R)}\,, \label{eq::mu2}
\end{align}
where $P(\bfs_{i+1}|\bfs_i)$ evaluates to $1$ on a tree branch if $i$ is a parity level and $1/2$ otherwise, if all information symbols are equiprobable. The second term in the summation can be computed recursively using a lattice \cite{bahlOptimalDecodingLinear1974, buttigiegImprovedBitError2015}. Note that the final two terms in~(\ref{eq::mu2}) are of the form $\sum_{\bfx \in \{0,1\}^{N'}} P(\bfx, \bfy_1^{R'})$, which signifies the probability of receiving a specific vector of length $R'$ given that $N'$ symbols were transmitted. If the information bits are uniformly distributed, both of these terms can be computed via the lattice-based procedure stated in \cite[Sec.~III-A]{banerjeeSequentialDecodingMultiple2024}. Otherwise, the codeword bit probabilities $P(x_i)$ should not be assumed to be uniform and can be computed by performing a forward pass on the syndrome trellis, i.e., using $P(x_i = a) = \sum_{(\bfs_i, \bfs_{i+1}) : x_i = a} P(\bfs_i)P(\bfs_{i+1}, x_i|\bfs_i)$ where $a \in \{0,1\}$ and $P(\bfs_{i+1}, x_i|\bfs_i)=P(\bfs_{i+1}|\bfs_i)$. 

In case of multiple received sequences $\bfy_1, \ldots, \bfy_M$, as in \cite{banerjeeSequentialDecodingMultiple2024}, we consider the augmented tree mentioned in Section~\ref{subsec::channel-code-tree}, so that a specific tree node, say $\bfv_0^t$, now refers to the syndrome states $\bfs_0^t = (\bfs_0, \ldots, \bfs_t)$ and the sequence of drift state vectors $(\bfd_0, \ldots, \bfd_t)$, where $\bfd_i=(d_{i,1}, \ldots, d_{i,M})$ denotes the drift of each of the $M$ received sequences after the transmission of $i$ bits. Accordingly, $\mu(\bfv_0^t)$ transforms into
\begin{align}
    &\mu(\bfv_0^t) \! = \log P(\bfv_0^t, \bfy_1, \ldots, \bfy_M) - \log P(\bfy_1, \ldots, \bfy_M) \nonumber \\
    &\approx \! \sum_{i=0}^{t-1 } \! \Big( \sum_{j=1}^{M}\log P((\bfy_j)_{i+d_{i,j}+1}^{i+1+d_{i+1,j}}, d_{i+1,j}|\bfs_i^{i+1},d_{i,j}) \nonumber \\
    & + \! \log P(\bfs_{i+1}|\bfs_i)   \!\Big) \! + \! \sum_{j=1}^{M} \log \frac{\sum_{\widetilde{\bfs}_{t+1}^{N}} P((\bfy_j)_{t+d_{t,j}+1}^{R_j}, \widetilde{\bfs}_{t+1}^{N}| d_{t,j})}{P((\bfy_j)_1^{R_j})}\,, \nonumber
\end{align}
where $R_j$ denotes the length of the $j$th received sequence $\bfy_j$.

\subsection{Bidirectional Decoding}

The principle of bidirectional decoding \cite{kallelBidirectionalSequentialDecoding1997} is to employ a \emph{forward decoder} and a \emph{backward decoder} that start from the two extreme ends of the joint channel and code tree, and progress toward each other as each seeks to find the path with the highest metric along their respective directions. It is worth noting that while the backward decoder works on the same tree as its forward counterpart, the drift state vector of its initial node consists of the net drift of each received sequence, and its reverse operating direction implies that the information and parity levels may be defined differently. The algorithm terminates when either both the forward and backward decoders meet at a common node in the tree or one of the decoders reaches a terminal node. For more details, we refer the reader to \cite{kallelBidirectionalSequentialDecoding1997}.

\section{Numerical Results}

\begin{table}[t!]
	\vspace{-1mm}
	\setlength{\tabcolsep}{2pt}
	\centering
	\caption{Convolutional Codes for Simulations}
	\renewcommand{\arraystretch}{1.1}
	\begin{tabular}{clcc} \specialrule{1.2pt}{0pt}{0pt}
		Code & $[n,k]$ & Parity-check matrix & $d_{\mathrm{free}}$ \\ \specialrule{.8pt}{0pt}{0pt}
		CC1 & $[10,7]$ & $(5,1,62, 114, 22, 214, 158, 169, 241, 177)$ & $6$\\
		CC2 & $[11,9]$ & $(116,87,115,26,93,15,109,75,107,205,167)$ & $5$ \\
		\specialrule{1.2pt}{0pt}{4pt}
	\end{tabular}
	\label{tab::codes}
\end{table}

\begin{figure}[t] 
	\centering
	\begin{tikzpicture}
		\begin{axis}[legend style={nodes={scale=0.65, transform shape}},
			xmode=log, ymode=log,
			xmin=2e-3, xmax=0.1001,
			ymin=1.5e-6, ymax=0.02,
			xlabel=\textsc{$\pd$}, ylabel={BER},
			legend pos=north west, legend style={font=\footnotesize},
			grid=both]
            \addplot plot[solid, stack, mark=diamond*] coordinates {
                (0.004, 0.000009)
                (0.008, 0.000078)
                (0.01, 0.00017425925925925926)
                (0.015, 0.00091)
                (0.02, 0.002764074074074074)
                (0.025, 0.006728703703703704)
            };
			\addlegendentry{CC2 Stack}
            \addplot plot[solid, bistack, mark=diamond*] coordinates {
                (0.004, 0.000015)
                (0.008, 0.000058)
                (0.01, 0.000083)
                (0.015, 0.0002825925925925926)
                (0.02, 0.0006174074074074074)
                (0.025, 0.0016996296296296297)
			};
			\addlegendentry{CC2 Bi-Stack}
            \addplot plot[solid, bcjr, mark=diamond*] coordinates {
                (0.004, 6.944e-6)
				(0.008, 0.000106018)
				(0.01, 0.000239)
				(0.015, 0.0010388)
				(0.02, 0.00319)
			};
			\addlegendentry{CC2 Sep.-BCJR}
            \addplot plot[dash dot, stack, mark=diamond*] coordinates {
                (0.004, 4.380952380952381e-06)
                (0.008, 3.847619047619047e-05)
                (0.01, 7.733333333333333e-05)
                (0.015, 0.000204)
                (0.02, 0.0006685714285714285)
                (0.025, 0.001525170068027211)
                (0.03, 0.003590952380952381)
			};
			\addlegendentry{CC1 Stack}
            \addplot plot[dash dot, bistack, mark=diamond*] coordinates {
                (0.008, 1.9047619047619046e-05)
                (0.01, 4.857142857142857e-05)
                (0.015, 0.00011295238095238095)
                (0.02, 0.00021523809523809524)
                (0.025, 0.00039197278911564623)
                (0.03, 0.0007295238095238096)
			};
			\addlegendentry{CC1 Bi-Stack}
            \addplot plot[dash dot, bcjr, mark=diamond*] coordinates {
                (0.004, 2.1428571428571427e-06)
                (0.008, 1.595238095238095e-05)
                (0.01, 3.476190476190476e-05)
                (0.015, 0.00011333333333333333)
                (0.02, 0.00031666666666666665)
                (0.025, 0.0007898412698412699)
                (0.03, 0.0017447619047619048)
			};
			\addlegendentry{CC1 Sep.-BCJR}
		\end{axis}
	\end{tikzpicture}
 \vspace{-2ex}
	\caption{BER versus $\pd$ when $\pin=\ps=0$ and $M=2$. CC1 and CC2 have lengths $N=126$ and $N=139$, respectively.}
	\label{fig::ber-del-only}
\end{figure}
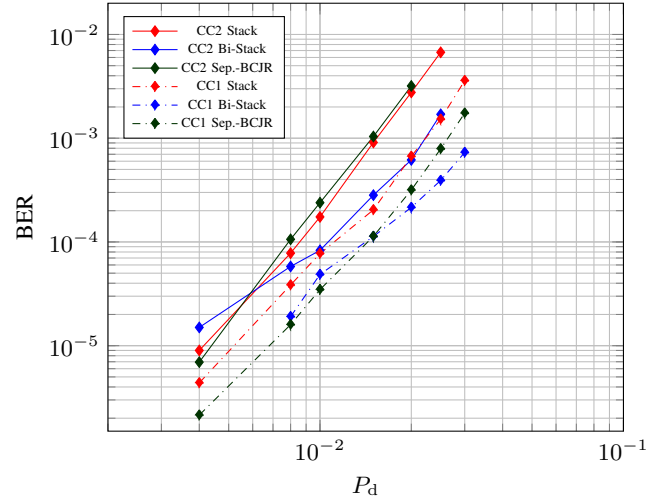

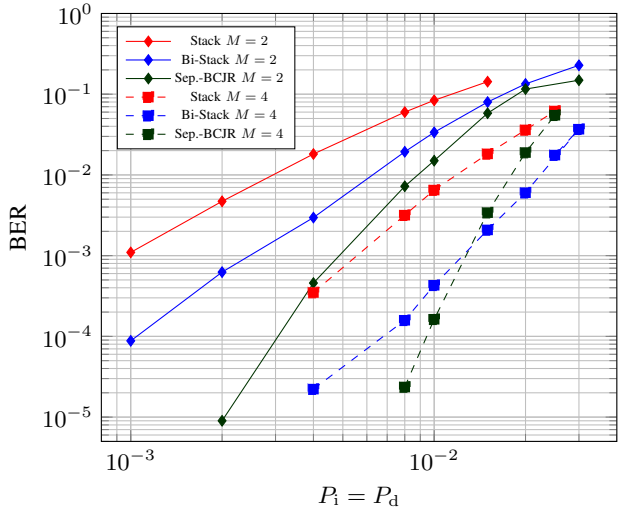
\begin{figure}[t] 
	\centering
	\begin{tikzpicture}
		\begin{axis}[legend style={nodes={scale=0.65, transform shape}},
			xmode=log, ymode=log,
			xmin=8e-4, xmax=0.04,
			ymin=5e-6, ymax=1.0,
			xlabel=\textsc{$\pin = \pd$}, ylabel={BER},
			legend pos=north west, legend style={font=\footnotesize},
			grid=both]
            \addplot plot[solid, stack, mark=diamond*] coordinates {
                (0.001, 0.0010975)
                (0.002, 0.004711) 
                (0.004, 0.018153) 
                (0.008, 0.0599765) 
                (0.01, 0.0839755) 
                (0.015, 0.14295809881175736)
			};
			\addlegendentry{Stack $M=2$}
            \addplot plot[solid, bistack, mark=diamond*] coordinates {
				(0.001, 8.8e-05)
                (0.002, 0.000625) 
                (0.004, 0.0029615) 
                (0.008, 0.019344) 
                (0.01, 0.0336935) 
                (0.015, 0.079944) 
                (0.02, 0.134027) 
                (0.03, 0.2280205)
			};
			\addlegendentry{Bi-Stack $M=2$}
			\addplot plot[solid, bcjr, mark=diamond*] coordinates {			
                (0.002, 9e-06)
		          (0.004, 0.000458)
		          (0.008, 0.0072285) 
		          (0.01, 0.0150045)
		          (0.015, 0.0581445) 
		          (0.02, 0.1157065)
		          (0.03, 0.149)
			};
			\addlegendentry{Sep.-BCJR $M=2$}
            \addplot plot[dashed, stack, mark=square*] coordinates {
                (0.004, 0.00034953703703703704)
                (0.008, 0.003172222222222222)
                (0.01, 0.006492161670195773)
                (0.015, 0.018227314814814814)
                (0.02, 0.03611435185185185)
                (0.025, 0.061849537037037036)
			};
			\addlegendentry{ Stack $M=4$}
            \addplot plot[dashed, bistack, mark=square*] coordinates {
                (0.004, 2.2222222222222223e-05)
                (0.008, 0.00015833333333333332)
                (0.01, 0.00042962962962962963)
                (0.015, 0.0020837962962962963)
                (0.02, 0.006022222222222222)
                (0.03, 0.03682916666666667)
                (0.025, 0.017604861436421364)
			};
			\addlegendentry{ Bi-Stack $M=4$}
			\addplot plot[dashed, bcjr, mark=square*] coordinates {		
                (0.008, 2.361111111111111e-05)  
                (0.01, 0.00016296296296296295)
                (0.015, 0.0034097222222222224)
                (0.02, 0.01892962962962963)
                (0.025, 0.05488148148148148)
			};
			\addlegendentry{Sep.-BCJR $M=4$}
		\end{axis}
	\end{tikzpicture}
 \vspace{-2ex}
	\caption{BER versus $\pin = \pd$ when $\ps=0$ for CC2 with varying number of received sequences, $M$. The codewords are of length $N=139$.}
	\label{fig::ber-11-9}
\end{figure}

    \begin{figure}[t]
		\centering
		\begin{tikzpicture}
			\begin{axis}[legend style={nodes={scale=0.65, transform shape}},
				xmode=log, ymode=log,
				xmin=8e-4, xmax=0.05,
				ymin=0.3, ymax=3e4,
				xlabel=\textsc{$\pin = \pd$}, ylabel={$\nu$},
				legend pos=south west, legend style={font=\scriptsize},
				grid=both]
                \addplot plot[solid, bistack, mark=diamond*] coordinates {
					(0.001, 10533.711558903535) 
                    (0.002, 8761.115422608293)
                    (0.004, 4776.0784983486465)
                    (0.008, 1037.1684689018073)
                    (0.01, 598.3822786036322)
                    (0.015, 239.31850312832736)
                    (0.02, 149.03694717787593) 
                    (0.03, 68.66417722321819)
				};
				\addlegendentry{Bi-Stack $M=2$}
                \addplot plot[solid, stack, mark=diamond*] coordinates {
					(0.001, 1564.7673061494077) 
                    (0.002, 449.22445632604735) 
                    (0.004, 129.51652940060424)
                    (0.008, 42.40807551907821) 
                    (0.01, 30.916747643951542)
                    (0.015, 18.277082968299133)
				};
				\addlegendentry{Stack $M=2$}
                \addplot plot[dashed, bistack, mark=square*] coordinates {
                    (0.001, 151.2861425892022)
                    (0.002, 168.70736861982266)
                    (0.004, 199.35966736858938)
                    (0.008, 206.29581609275763)
                    (0.01, 143.25355169304518)
                    (0.015, 49.92731563088872)
                    (0.02, 19.065681225970266)
                    (0.03, 3.2345365024918835)
                    (0.025, 6.644407013159615)
				};
				\addlegendentry{Bi-Stack $M=4$}
               \addplot plot[dashed, stack, mark=square*] coordinates {
                    (0.001, 149.31636551187194)
                    (0.002, 141.657907714614)
                    (0.004, 69.14296190112292)
                    (0.008, 13.057885184453928)
                    (0.01, 6.45453979592506)
                    (0.015, 2.795124279126523)
                    (0.02, 1.5282485735171987)
                    (0.025, 0.9634463932826605)
				};
				\addlegendentry{Stack $M=4$}
			\end{axis}
		\end{tikzpicture}
  \vspace{-2ex}
		\caption{Complexity reduction $\nu$ versus $\pin=\pd$ when $\ps=0$ for CC2 with length $N=139$.}
		\label{fig::nu}
	\end{figure}
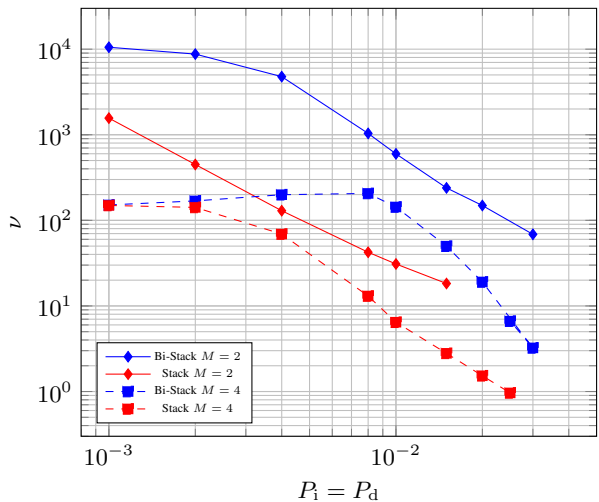
To evaluate the performance of the sequential and the separate-BCJR \cite{maaroufConcatenatedCodesMultiple2023} decoders over the syndrome trellis, we consider two convolutional codes of distinct rates from \cite{rosnesMaximumLengthConvolutional2004}, denoted by CC1 and CC2 and stated in Table~\ref{tab::codes}. We denote by $d_{\mathrm{free}}$ the free distance of a convolutional code. The parity-check matrix is represented as a vector of integers, wherein each integer is the decimal representation of a column of the polynomial parity-check matrix after expanding each polynomial into its coefficients. We simulated the transmission of terminated codewords of different lengths  (offset by a random sequence \cite{buttigiegImprovedBitError2015}) over channels with distinct insertion and deletion probabilities. To limit the decoding complexity, we ignore branches with net drift exceeding magnitude $1$, and set limits on the maximum and minimum overall drifts using the technique in \cite{briffa_timevarying_2014}. For the stack algorithm, we employ a stack of size $3\cdot 10^5$. If the number of forward steps exceeds $4\cdot 10^5$, we declare an erasure, pick the partial codeword corresponding to the topmost tree node in the stack, and randomly guess the remaining undecoded bits.

\subsection{Decoding Performance}

We compare the bit error rates (BERs) of the unidirectional and bidirectional stack decoding algorithms with those of the separate-BCJR decoder \cite{maaroufConcatenatedCodesMultiple2023}, which involves executing the standard BCJR algorithm for each of the $M$ received sequences $\bfy_j$ to obtain the a posteriori probabilities  $P(x_t | \bfy_j)$, and combining them as 
\begin{align*}
    P(x_t | \bfy_1, \ldots, \bfy_M) &\approx  \frac{\prod_{j=1}^{M} P(x_t | \bfy_j) }{ P(x_t)^{M-1} }.
\end{align*}

The simulations are performed over channels with $\pin = 0$ and with $\pin=\pd$, while $\ps = 0$ (Fig.~\ref{fig::ber-del-only} and Fig.~\ref{fig::ber-11-9}, respectively). We compare the BER performance of CC1 and CC2 for $M=2$ received sequences in Fig.~\ref{fig::ber-del-only}, and observe that for both codes, under low noise, the separate-BCJR decoder outperforms the unidirectional and  bidirectional stack algorithms since it uses the joint channel and code trellis to compare a received sequence with all possible codewords and sequences of drift vectors. In contrast, the stack algorithms only examine the codewords that seem promising and are consequently more error-prone, particularly for codes with smaller free distance. However, as the channel noise worsens, the BER for the separate-BCJR decoder rises faster compared to that of the bidirectional stack algorithm and is soon surpassed by the latter. This is because the bidirectional stack algorithm decodes multiple received sequences jointly, which assists significantly in the accurate estimation of the sequence of syndrome states. In essence, at higher noise levels, joint decoding of multiple sequences compensates for the BER loss caused by the sequential decoder's partial examination of the joint channel and code tree. Since a larger free distance offers a greater advantage to separate-BCJR decoding, the deletion probability where the bidirectional algorithm overtakes is higher for CC1. We also observe in Fig.~\ref{fig::ber-11-9} how the BER performance of these decoders varies with more received sequences for CC2. For $M=4$, the bidirectional stack algorithm surpasses the performance of the separate-BCJR decoder at elevated noise levels, while this advantage is not observed when $M=2$. Additionally, the unidirectional stack algorithm performs consistently worse than its bidirectional counterpart. This is attributed to its increased susceptibility to erasures under high noise levels. However, under low noise levels, this behavior differs from that observed on the binary symmetric channel \cite{kallelBidirectionalSequentialDecoding1997}. We believe this might be due to the inherent challenges imposed by the infinite memory of the insertion and deletion channel. While the estimation of the drift sequence becomes more difficult further into the codeword, this challenge is eased for the bidirectional stack algorithm since, in the best case, each of its component decoders needs to predict only half of the complete sequence of drift vectors.



\subsection{Simulated Complexity}

We compare the computational complexity of the unidirectional and bidirectional stack decoders and the separate-BCJR decoder over the syndrome trellis in terms of the number of metric computations performed. For the separate-BCJR decoder, this quantity equals the number of received sequences $M$ times the number of branches in the trellis for a single received sequence, denoted by $\cB_{\mathrm{tr}}$. In contrast, the complexity of a stack decoder is  the average number of nodes visited (Step~2, Section~\ref{subsec::seq-dec}), represented as $\cF_{\mathrm{av}}$, which is exponential in $M$, as hinted in Section~\ref{subsec::channel-code-tree}. Hence, similarly to \cite{banerjeeSequentialDecodingMultiple2024}, we define the complexity reduction factor as
\begin{align*}
    \nu &= \frac{\text{Complexity of separate-BCJR decoding}}{\text{Avg. complexity of stack algorithm}} = \frac{M \cB_{\mathrm{tr}} }{\cF_{\mathrm{av}}}\,.
\end{align*}

Fig.~\ref{fig::nu} demonstrates significant reductions in decoding effort offered by both stack algorithms when compared with the separate-BCJR decoder. An initial rise in $\nu$ may occur because the complexity of the separate-BCJR decoder increases with higher $\pin = \pd$, due to the larger number of drift states to be considered, while the complexity of the unidirectional and bidirectional stack decoders remains fairly constant under low noise levels. The bidirectional stack algorithm reduces complexity even further compared to its unidirectional counterpart for $M=2$, a trend also seen in earlier work \cite{kallelBidirectionalSequentialDecoding1997, xuBidirectionalFanoAlgorithm2009}. The reason lies in the fact that the unidirectional stack algorithm, upon encountering an uncorrectable pattern, will examine more incorrect paths until it finds the true path in the joint channel and code tree, while the bidirectional decoder attacks this erroneous region of the received sequence from two directions \cite{kallelBidirectionalSequentialDecoding1997}. In this manner, the latter may reduce the amount of computations in such a noisy region by at most a factor of two. With the availability of more received sequences, these events become less likely. However, beyond a certain error probability, the complexity of both stack decoders grows rapidly due to the increasing likelihood of erasures, and approaches the complexity of the separate-BCJR decoder. We also observe in Fig.~\ref{fig::nu} that the computational advantage offered by the sequential decoders fades as the number of received sequences increases, since the complexity of the separate-BCJR decoder increases linearly with the number of received sequences $M$, while for both stack decoders the complexity grows exponentially with $M$.

\balance
\printbibliography

\end{document}